\documentclass[twocolumn,superscriptaddress,prc,nofootinbib]{revtex4-1}

\usepackage{graphicx}
\usepackage[colorlinks = true,
            linkcolor = cyan,
            urlcolor  = blue,
            citecolor = red,
            anchorcolor = blue]{hyperref}\usepackage{color}
\usepackage{amssymb}
\usepackage{amsthm}
\usepackage{textcomp}
\usepackage{mathtools}
\usepackage{comment}
\usepackage[separate-uncertainty,retain-explicit-plus,per-mode = symbol]{siunitx}
\usepackage{url}

\newcommand{\krfive}{\mbox{$^{85}$Kr}}

\newcommand{\arforty}{\mbox{$^{40}$Ar}}
\newcommand{\poten}{\mbox{$^{210}$Po}}
\newcommand{\pofour}{\mbox{$^{214}$Po}}
\newcommand{\poeight}{\mbox{$^{218}$Po}}
\newcommand{\bifour}{\mbox{$^{214}$Bi}}
\newcommand{\biten}{\mbox{$^{210}$Bi}}
\newcommand{\bipo}{\mbox{$^{214}$Bi-$^{214}$Po}}
\newcommand{\pbfour}{\mbox{$^{214}$Pb}}
\newcommand{\pbten}{\mbox{$^{210}$Pb}}
\newcommand{\pbsix}{\mbox{$^{206}$Pb}}
\newcommand{\rntwo}{\mbox{$^{222}$Rn}}
\newcommand{\amone}{\mbox{$^{241}$Am}}
\newcommand{\mgf}{\mbox{MgF$_{\text{2}}$}}

\newcommand{\keVee}{\mbox{keV$_{\text{ee}}$}}
\newcommand{\fp}{\mbox{F$_{\text{prompt}}$}}

\newcommand{\mgcm}{\mbox{\text{mg/cm}$^{2}$}}

\begin{document}

\title{First measurement of surface nuclear recoil background for argon dark matter searches}

\author{Jingke Xu}
\email[Corresponding author, ] {xu12@llnl.gov}
\altaffiliation{Current Address: Lawrence Livermore National Laboratory, Livermore, California 94550, USA}
\affiliation{Department of Physics, Princeton University, Princeton, New Jersey 08544, USA}
\author{Chris Stanford}
\thanks{Jingke Xu and Chris Stanford contributed equally to this work.}
\affiliation{Department of Physics, Princeton University, Princeton, New Jersey 08544, USA}
\author{Shawn Westerdale}
\affiliation{Department of Physics, Princeton University, Princeton, New Jersey 08544, USA}
\author{Frank Calaprice}
\affiliation{Department of Physics, Princeton University, Princeton, New Jersey 08544, USA}
\author{Alexander Wright}
\affiliation{Department of Physics, Engineering Physics, and Astronomy, Queen's University, Kingston, Ontario, Canada. }
\author{Zhiming Shi}
\altaffiliation{Current Address: Google Inc., Mountain View, California 94043, USA}
\affiliation{Department of Physics, Princeton University, Princeton, New Jersey 08544, USA}

\date{\today}

\begin{abstract}

One major background in direct searches for weakly interacting massive particles (WIMPs) comes from the deposition of radon progeny on detector surfaces. 
The most dangerous surface background is the \pbsix\ recoils produced by \poten\ decays. 
In this letter, we report the first characterization of this background in liquid argon. 
The scintillation signal of low energy Pb recoils is measured to be highly quenched in argon, and 
we estimate that the 103\,keV \pbsix\ recoil background will produce a signal equal to that of a $\sim$5\,keV (30\,keV) electron recoil (\arforty\ recoil). 
In addition, we demonstrate that this dangerous \poten\ surface background can be suppressed, using pulse shape discrimination methods,  by a factor of $\sim$100 or higher, which can make argon dark matter detectors near background-free and enhance their potential for discovery of medium- and high-mass WIMPs. 
We also discuss the impact on other low background experiments.

\end{abstract}


\maketitle



 
Noble liquid detectors have demonstrated exceptional sensitivity in direct searches for weakly interacting massive particles (WIMPs), a popular candidate for dark matter. Over the past decade, xenon-based experiments including XENON100~\cite{XENON2012_225dResult} and LUX~\cite{LUX2015_Reanalysis} have achieved the highest sensitivities in this field. Recently, argon-based experiments have also developed key technologies necessary for sensitive WIMP searches, as demonstrated by the DarkSide-50 experiment~\cite{DarkSide_AAr_2015,DarkSide_UAr_2015}.
The DEAP-3600 experiment~\cite{Boulay_deap-3600_2012}, which is being commissioned at SNOLAB, is expected to achieve a higher dark matter sensitivity than that of current xenon experiments, especially for high-mass WIMPs. 
With the powerful pulse shape discrimination (PSD) capability of argon, 
argon dark matter searches at multi-tonne scales can be free of electron recoil backgrounds from solar neutrinos~\cite{Strigari2014_NeutrinoBg} and from radioactive decays of radon progeny and \krfive~\cite{XENON1T_2016}, all of which compromise the sensitivity of xenon experiments.
 
Due to the low expected interaction rate between WIMP dark matter and ordinary matter, it is critical for WIMP search experiments to achieve a very low background rate, especially for nuclear recoil background that can mimic a WIMP interaction. 
One such nuclear recoil background can result from the exposure of detector surfaces to radon that is naturally present in the environment, specifically in air and in ground water. 
Through the following decay sequence,
\begin{eqnarray}
\rntwo &\rightarrow& \poeight \rightarrow \pbfour \rightarrow \bifour \rightarrow \pofour \nonumber \\
 &\rightarrow& \pbten \rightarrow \biten \rightarrow \poten \rightarrow \pbsix  \nonumber
\end{eqnarray}
\poten\ and other radon progeny can be produced and become attached to detector surfaces. 
The decay of \poten\ produces a \pbsix\ recoil and an $\alpha$ particle. 
\begin{equation}
\poten \xrightarrow{138\,d}  \pbsix\ (103\,\text{keV})  + \alpha\ (5.3\,\text{MeV}) \nonumber
\end{equation}
The recoiling \pbsix\ nucleus can make a dangerous background if it is recorded in the active volume of the detector. 
Unlike the short-lived nuclides in the early radon chain, \pbsix\ could be produced many years after radon exposure owing to the long half-life ($\sim$22 years) of  \pbten.

\pbsix\ recoils have been identified as one of the most important backgrounds in several major dark matter experiments, such as LUX~\cite{LUX2015_Reanalysis}, SuperCDMS~\cite{CDMS2015_Ge,Redl2014_Pb206CDMS}, CoGeNT~\cite{CoGeNT_2013}, and CRESSTII~\cite{Cresst_2014}.
For example, a surface \poten\ $\alpha$-decay rate of $\sim$ 35 mHz was detected in the LUX experiment~\cite{lux_lee_phd}, 
and \pbsix\ was the major remaining background observed in the WIMP search signal region even after position cuts~\cite{LUX2015_Reanalysis}.
For argon-based dark matter experiments like DarkSide-50 and DEAP-3600, position sensitivity as accurate as that in xenon detectors has not been achieved, which makes these experiments more vulnerable to surface background contamination or large loss of fiducial mass. 
Indeed, the \pbsix\ problem has been considered the most dangerous background for the DEAP-3600 experiment~\cite{DEAP2013_SurfBG}.

This paper presents a study of the surface \pbsix\  nuclear recoil background for argon dark matter experiments. 
We will report the first scintillation measurement of low energy Pb recoils in liquid argon, 
and then characterize the full surface background by taking into consideration the signals induced by the $\alpha$ particles accompanying the Pb recoils. 
Because argon experiments usually use a wavelength shifter (WLS) coating on the interior detector surfaces for the detection of vacuum ultraviolet (VUV) argon scintillation light, the $\alpha$ particles can produce additional scintillation photons in the WLS.
With this scintillation signal, we demonstrate that this surface background in argon dark matter experiments can be significantly suppressed.
The impact of this research on DarkSide-50, DEAP-3600 and other low-background experiments will be discussed.

Although the \pbsix\ recoils from \poten\ decays are produced at a relatively high energy (103\,keV), most of the energy is non-radiatively dissipated as heat and cannot be detected in argon detectors. 
Due to the low detectable energy and the lack of signal tagging methods, no characterization of this \pbsix\ recoil signal has been reported. 
In this work, we took an alternative approach, namely, to investigate the \pbsix\ recoil signal by studying the \pbten\ recoils in the $\alpha$ decays of \pofour. 
\pofour\ can be produced by the $\beta$ decay of \bifour\ in the \rntwo\ chain; the short half life ($\sim$164 $\mu$s) of \pofour\ means that the \bifour\ and \pofour\ decays will be in delayed coincidence, and this \bipo\  coincidence can be used to efficiently tag \pofour\ decays. \pofour\ decays through the reaction
\begin{equation}
\pofour \xrightarrow{164\,\mu s} \pbten\ (146\,\text{keV}) + \alpha\ (7.7\,\text{MeV}) \nonumber
\end{equation}
The slightly higher \pbten\ recoil energy makes the direct measurement more viable; at the same time, one does not expect the scintillation efficiency of argon for \pbsix\ recoils to differ significantly from that for \pbten. 

\begin{figure}[h!]
\centering
\includegraphics[height=.55\textwidth]{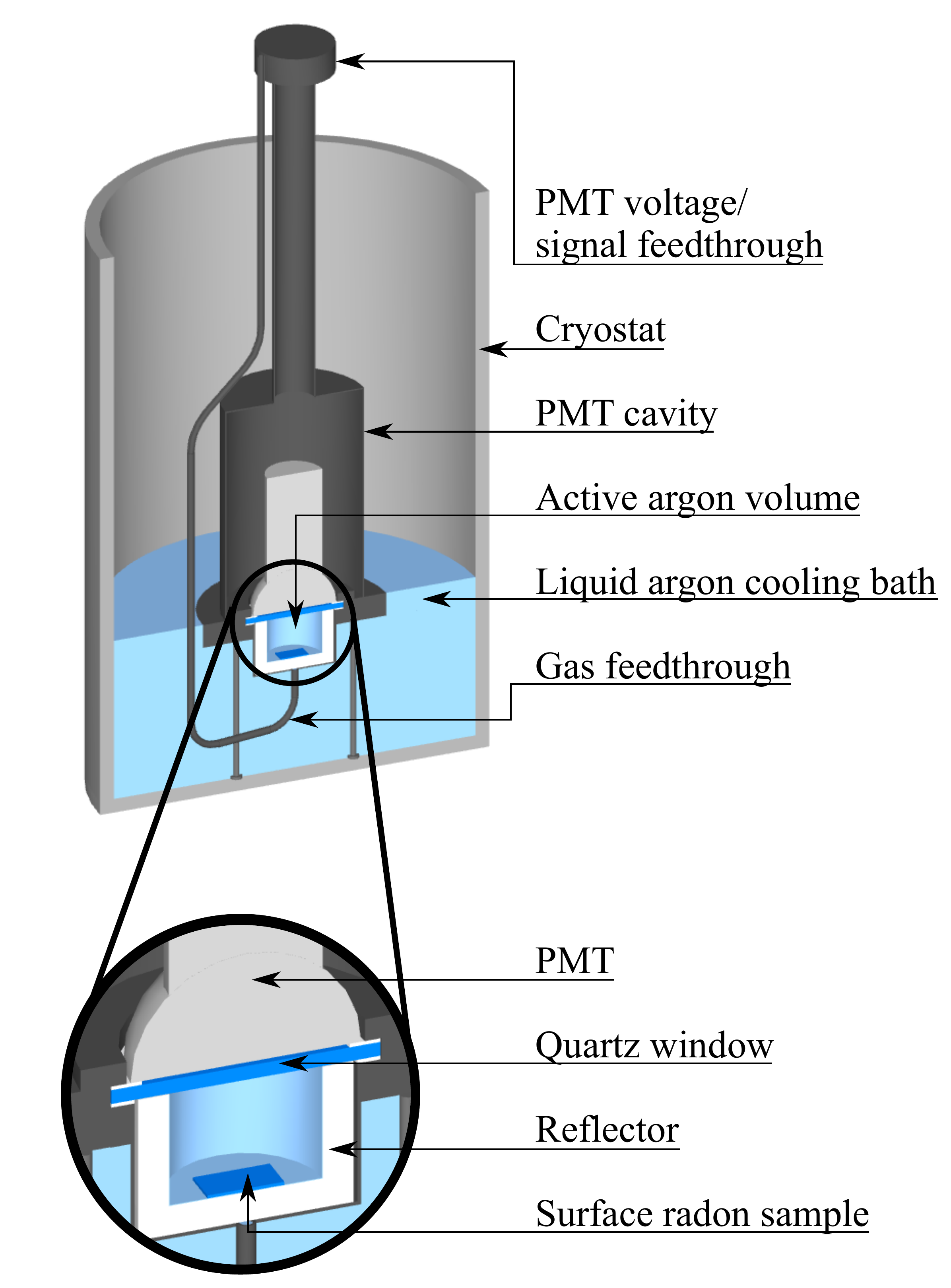}
\caption{Illustration of the single-phase liquid argon detector used in this study. The lower chamber ($\Phi$72\,mm$\times$48\,mm) hosts a Spectralon reflector cell that contains the radon sample in purified liquid argon; the upper chamber hosts a Hamamatsu R11065 photmultiplier tube (PMT). The two chambers are hermetically sealed with a quartz window. The reflector cell and the quartz window are coated with a WLS for argon light collection. An external liquid argon bath provides the needed cooling power.}
\label{fig:apparatus}
\end{figure}

In this experiment, we first collected \rntwo\ progeny onto a VUV-reflective mirror ($>$85\% reflectivity for argon scintillation light at 128\,nm~\cite{ActonAlMgF}) by exposing its reflective side to a radon-argon gas mixture with a \rntwo\ activity of $\sim$2\,MBq~\cite{xu_thesis}. The VUV mirror consists of a highly reflective aluminum coating on a quartz substrate and a 25$\pm$10\,nm \mgf\ protective layer to prevent the aluminum from oxidizing. 
During the exposure, the progeny of \rntwo\ plated out on the VUV mirror and \pbfour\ quickly accumlated due to its relatively long half life (27\,min). 
After about 3 hours, this VUV mirror was removed from the radon collection chamber and deployed into a liquid argon detector as illustrated in Fig.~\ref{fig:apparatus}. The detector was then pumped and purged for several cycles to remove electronegative impurities, cooled down to 87\,K with an external liquid argon bath, and filled with purified argon for scintillation measurements. 

As \pofour\  decayed on the surface of the VUV mirror, the daughter \pbten\ nucleus could recoil into the liquid argon and produce VUV argon scintillation light. 
Meanwhile, the accompanying $\alpha$ particle will lose its energy in the mirror and not produce significant light due to the thinness of \mgf. 
The \pbten\ light was efficiently collected by a photomultiplier (PMT) through the use of the VUV mirror and a WLS coating on the Spectralon reflector. 
Due to the 27 min lifetime of \pbfour, the \bipo\ coincidence rate became negligible within several hours of the initial radon exposure. The whole measurement therefore had to be completed within 6--8 hours. 
This was achieved by the specially designed single-phase liquid argon detector illustrated in Fig.~\ref{fig:apparatus}. 

\begin{figure}[h!]
\centering
\includegraphics[width=.48\textwidth]{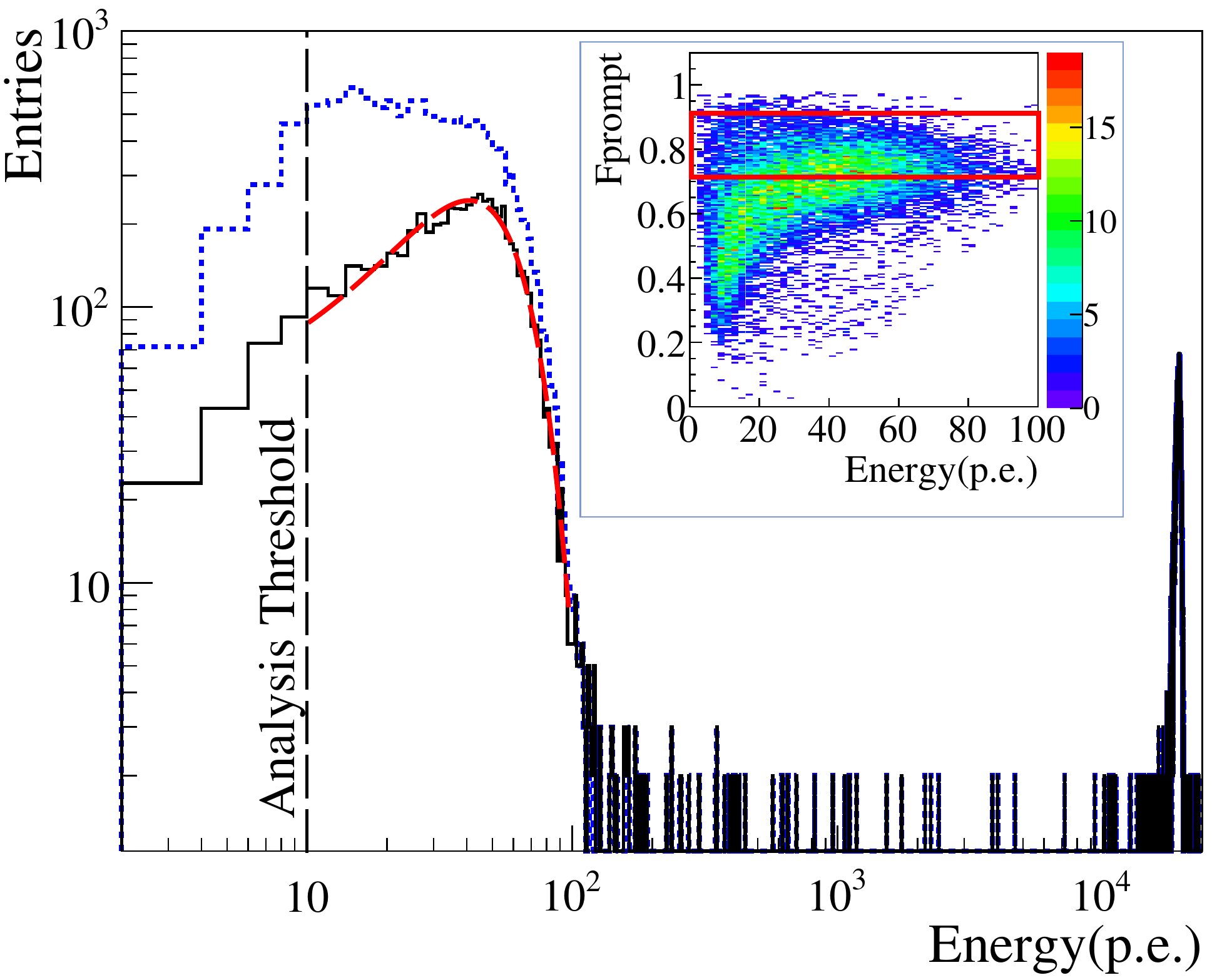}
\caption{Scintillation spectra of liquid argon excited by \pofour\ decays with (solid black) and without (dotted blue) \fp\ cut. 
The  \pbten\ recoils are observed below 100\,p.e. and the $\alpha$ events are observed around 20,000\,p.e.. 
The inset figure shows the \fp\ distribution of \pbten\ recoils. 
The \fp\ cut selected events within 2$\sigma$ above the most probable \fp\ value, and a Gaussian fit to the spectrum is also shown.}
\label{fig:po214_espectrum}
\end{figure}

The energy spectrum of the \pofour\ decay events, identified by the \bipo\ coincidence, is shown in Fig.~\ref{fig:po214_espectrum} (dotted blue). The energy scale is given by the number of photoelectrons (p.e.) detected by the PMT, which is proportional to the number of scintillation photons produced.
Due to the presence of electronegative impurities in the liquid argon, a  small fraction of the triplet argon scintillation was lost; a correction for this effect was made based on the measured triplet scintillation lifetime and singlet-to-triplet ratio.  
Two groups of events were observed in the \pofour\ decays. The high energy events around 20,000\,p.e. were easily identified as $\alpha$ particles and the low energy events below 100\,p.e. can be attributed to \pbten\ recoil nuclei. The rates of the two event groups are approximately equal, which confirmed the explanation of their origins. 

However, the \pbten\ energy spectrum exhibited a broad distribution with a high-energy cutoff instead of a mono-energetic peak. This could be explained by \pofour\ precursors recoiling into the surface and becoming embedded beneath it. When these embedded \pofour\ nuclei decayed the energy of the recoiling \pbten\ nuclei would be degraded by the time they reached the argon. 
The pulse shape parameter \fp, defined as the fraction of scintillation within the first 90\,ns, also exhibited lower values at lower energy for the embedded events, as shown in the inset of Fig.~\ref{fig:po214_espectrum}. This trend is similar to that of \arforty\ recoils measured by \cite{Cao:2015ks}, but occurs at higher energies for \pbten\ recoils. This behavior was qualitatively confirmed by SRIM simulations~\cite{ziegler_srim_2010} that predict that \pbten\ nuclei lose most of their energy to recoils with argon nuclei, which then scintillate. 
A single \pbten\ nucleus recoiling in argon will therefore produce a signal similar to that of multiple lower energy argon nuclei recoiling, explaining why \fp\ decreases at higher energies for \pbten\ recoils than it does for direct \arforty\ recoils.

To calculate the argon scintillation light yield for full energy \pbten\ recoils (146\,keV), we selected the \pbten\ events within the \fp\ range of 0.72 - 0.9, as shown in the inset of Fig.~\ref{fig:po214_espectrum}. 
These \pbten\ events were selected because they exhibited \fp\ values above the most probable value (0.72) and were only weakly affected by the embedded events. 
The resulting spectrum exhibited a Gaussian peak, 
the peak position of which was found at 41.2$\pm$0.6 p.e. using a $\chi^2$ fit. 
The uncertainty resulting from the \fp\ cut was evaluated by varying the cut range. 
The data acquisition threshold in the measurement was set to be $\sim$1.5 p.e., and the trigger efficiency is expected to be 100\% above the analysis threshold of 10 p.e.
The full energy \pbten\ light output was corrected to 45.7$\pm$2.3 p.e. after considering the VUV reflectivity of the aluminum mirror. A second measurement with a non-VUV-reflective silver foil as the radon backing material yielded consistent results after corrections.

Using the measured scintillation light yield of 6.2$\pm$0.2\,p.e./keV for 59.5 keV \amone\ $\gamma$s, we can obtain an electron-equivalent energy of 7.4$\pm$0.4 \keVee\ for the 146\,keV \pbten\ recoils. This result indicates a scintillation quenching factor of 19.7$\pm$1.2 relative to gammas/electrons and 6.2$\pm$0.5 relative to \arforty\ nuclear recoils, 
extrapolated from \cite{Scene_collaboration_observation_2013}. Simulations using the SRIM software~\cite{ziegler_srim_2010} indicate that the stopping power of \pbten\ recoils in liquid argon is $\sim$5 times higher than that of \arforty\ recoils. This high stopping power can cause 1) a high fraction of the \pbten\ energy to be dissipated as heat, decreasing the Lindhard factor, and 2) a high argon dimer decay rate through non-radiative channels, strengthening the Birks saturation. The SRIM simulations, however, under-predict the amount of quenching, possibly due to Lindhard model breaking down for Pb recoils at low energies, as suggested in~\cite{sorensen_atomic_2015}. 

Assuming that the scintillation efficiency of liquid argon is the same for \pbten\  recoils (146\,keV), measured in this work, and for \pbsix\ recoils (103 keV) produced by  \poten, the \pbsix\ surface background would produce only a modest $\sim$5 \keVee\ signal, similar to a $\sim$30\,keV \arforty\ recoil. This background is below the energy threshold of current argon dark matter experiments~\cite{DarkSide_AAr_2015}. 
However, since argon detectors usually use WLS coatings on the inner detector surfaces, 
when a surface \pbsix\ nucleus enters liquid argon, the $\alpha$ particle will enter the WLS and produce additional scintillation light. 
This $\alpha$ signal will contribute to the overall scintillation light, and make the \pbsix\ recoils more likely to become a background in argon-based dark matter experiments.

Therefore, we carried out a direct, {\it in situ} surface background measurement that combined scintillation signals from both the Pb recoils and the accompanying $\alpha$ particles. 
The measurement used a similar technique to the \pbten\ recoil experiment described earlier, but instead of leaving the $\alpha$ particles undetected, we used a WLS coating to produce a scintillation signal upon $\alpha$ particle incidence, as occurs for real surface backgrounds in argon detectors.  Again, we used \pofour\ decays to produce the signals and used the \bipo\ coincidence method to tag the surface \pofour\ events. 
In this study, two WLS chemicals were investigated: tetraphenyl-butadiene (TPB), the most widely used WLS in argon dark matter experiments, and 1,4-Diphenylbenzene (pTP, or p-Terphenyl). 
To ensure consistency in the comparison, we applied approximately the same coating thickness for both WLSs ($\sim$0.3-0.4 \mgcm), which was chosen to match that used in dark matter experiments. 

\begin{figure}[h!]
\centering
\includegraphics[width=.48\textwidth]{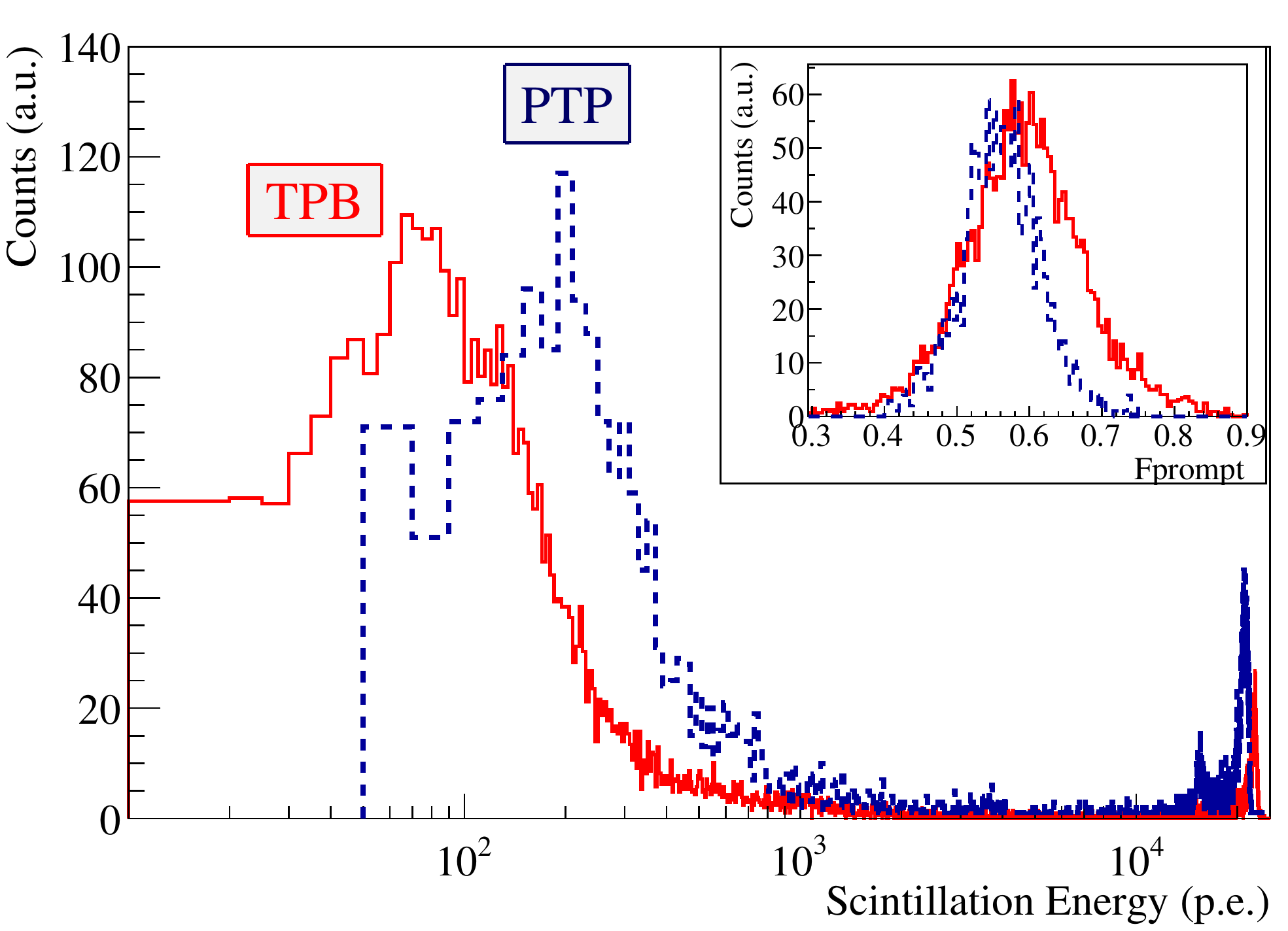}
\caption{The scintillation energy spectra of \pofour-induced surface events measured in argon, using TPB ($\sim$0.4 \mgcm) and pTP ($\sim$0.3 \mgcm) as the WLS, respectively. 
The low energy events ($\sim$100\,p.e.) contain the \pbten\ signals in argon and the $\alpha$ signals in WLS; the high energy events ($\sim$20,000\,p.e.) are dominated by $\alpha$ signals in argon. 
The inset shows the \fp\ distribution (15$\mu$s maximum integral window) for the low energy surface events.}
\label{fig:sb_insitu}
\end{figure}

The energy spectra of \pofour\ decay products, selected with the \bipo\ coincidence, are shown in Fig.~\ref{fig:sb_insitu}. 
As expected, two peaks are observed in each spectrum. 
The higher energy peak contains the full energy $\alpha$ signals in argon and the Pb recoils in the WLS.  
The lower energy peak contains the \pbten\ recoil signals in argon together with the $\alpha$-induced signals in the WLS.
Due to the additional WLS scintillation, the light output of the low energy peak greatly increased in comparison to that in Fig.~\ref{fig:po214_espectrum}.
The different amount of increase observed with the two WLSs agrees well with studies of WLS scintillation properties under $\alpha$ excitation~\cite{Pollmann_scintillation_2011, DEAP2015_TPBCryo, xu_thesis, WLS_paper_2016}. 
Due to the mix of argon scintillation and WLS scintillation in this measurement, the argon scintillation loss due to impurities is not corrected, but the effect  on the overall energy scale is estimated to be less than 5\%. 

The peak signals from the low energy surface background events were observed at 88\,p.e. and 192\,p.e. for TPB and pTP, respectively. Both fall in the dark matter search window as used in DarkSide-50~\cite{DarkSide_AAr_2015, DarkSide_UAr_2015}. 
The fact that this background has not been observed in DarkSide-50 could be partially explained
 by the relatively low statistics, 
and more importantly by the low ionization collection efficiency on the detector surfaces, which cause surface events to fail the analysis cuts, similar to that observed in LUX~\cite{lux_lee_phd}.  
Single phase argon dark matter experiments like DEAP-3600, on the other hand, only collect the scintillation signals and are therefore more vulnerable to surface background contamination~\cite{DEAP2013_SurfBG}. 

The surface background, however, can be suppressed using the pulse shape discrimination (PSD) method. The inset of Fig.~\ref{fig:sb_insitu} shows the overall \fp\ distribution of the low energy surface background events. 
 Owing to the relatively slow WLS scintillation under $\alpha$ excitation~\cite{WLS_paper_2016, DEAP2015_TPBCryo}, the overall \fp\ distribution of surface background events is strongly biased towards lower values than those of pure  nuclear recoils in liquid argon. 
Using the simple \fp\ PSD method, we estimate that the surface background can be suppressed by a factor of $\sim$10 (100) for TPB (pTP) at 50\% (90\%) \arforty\ recoil acceptance 
for coatings of the investigated thicknesses. 
Although TPB exhibits a low background rejection power using the simple \fp\ method, 
 a newly discovered long decay component in TPB under $\alpha$ excitation~\cite{DEAP2015_TPBCryo}  could in principle increase the PSD rejection to a factor of 100 or better~\cite{WLS_paper_2016}. 
As for pTP, the rejection power against the Pb recoil background can also be further improved by increasing the coating thickness, which will lower both the central value and the spread of the \fp\ distribution for the combined surface background events. 
Optimization of the surface background rejection power for argon dark matter experiments is beyond the scope of this paper; interested readers can find more on this topic in references \cite{xu_thesis} and \cite{WLS_paper_2016}. 

We point out that the surface background suppression power presented here is conservative for two reasons. First, we left out the correction for the loss of argon triplet scintillation due to impurities, and such a correction will lower the overall \fp\ values for the surface background events. Second, the \fp\ value of \pbsix\ recoils will be lower than that of \pbten\ due to the lower recoil energy, especially for those embedded under the detector surfaces. Both factors will increase the separation of the surface background from the nuclear recoil values in \fp\ distributions and enable stronger background rejection. 
A full evaluation of the \poten\ background in argon dark matter experiments requires extrapolation from the measured \pofour\ results, but we expect the background suppression factor to be at the same order of magnitude due to the relatively small contribution of the Pb recoils to this background. 

Finally, we comment that this method of rejecting surface backgrounds by detecting the $\alpha$ particles has potential applications beyond argon-based dark matter experiments. For example, coating the reflector surfaces of xenon experiments with a thin layer of \mgf\ or LiF, which can produce significant scintillation~\cite{Mikhailik2010_MgF, Baldacchini2005_LiF} under $\alpha$ excitations and which are transparent to xenon scintillation light, can help reject surface backgrounds reject surface backgrounds like those observed in LUX~\cite{LUX2013_FirstResult, lux_lee_phd}. For double-beta decay experiments like CUORE, the dominant surface background arises from $\alpha$ particles with partial energy deposition in the crystals~\cite{CUORE2015_0}. Similarly, the $\alpha$ particles may be detected with a thin coating of scintillating material on the surfaces of the crystals/supporting structures, which would allow this background to be suppressed. 

We thank Ben Loer for developing the data acquisition software that was used in this experiment. 
We are grateful to Peter Meyers for his insightful suggestions on both the measurements and the analyses. 
We thank Dongming Mei for discussions on the Pb recoil physics and thank Adam Bernstein and Brian Lenardo for reading through the manuscript. 
This work was supported by the NSF grants PHY0704220 and PHY0957083.
JX is an employee of the Lawrence Livermore National Laboratory. 
LLNL is operated by Lawrence Livermore National Security, LLC, for the U.S. Department of Energy, National Nuclear Security Administration under Contract DE-AC52-07NA27344.

\bibliographystyle{apsrev}
\bibliography{biblio}

\end{document}